\documentclass[10pt,conference]{IEEEtran}

\IEEEoverridecommandlockouts
\usepackage{cite}
\usepackage{url}
\usepackage{amsmath,amssymb,amsfonts}
\usepackage{algorithmic}
\usepackage{graphicx}
\usepackage{textcomp}
\usepackage{xcolor}
\def\BibTeX{{\rm B\kern-.05em{\sc i\kern-.025em b}\kern-.08em
    T\kern-.1667em\lower.7ex\hbox{E}\kern-.125emX}}
\begin{document}

\title{MSR Mining Challenge: The SmartSHARK Repository Mining Data\\
\thanks{This work was partially funded by DFG Grant 402774445.}
}

\author{\IEEEauthorblockN{1\textsuperscript{st} Alexander Trautsch}
\IEEEauthorblockA{\textit{Institute of Computer Science} \\
\textit{University of Goettingen}\\
Göttingen, Germany\\
alexander.trautsch@cs.uni-goettingen.de}
\and
\and
\IEEEauthorblockN{2\textsuperscript{nd} Fabian Trautsch}
\IEEEauthorblockA{ftrautsch@googlemail.com}
\and
\IEEEauthorblockN{3\textsuperscript{rd} Steffen Herbold}
\IEEEauthorblockA{\textit{Institute of Software and Systems Engineering} \\
\textit{TU Clausthal}\\
Clausthal-Zellerfeld, Germany\\
steffen.herbold@tu-clausthal.de}
}

\maketitle

\newcommand{\numberProjects}{77}
\newcommand{\numberCommits}{366,322}
\newcommand{\numberIssues}{163,057}
\newcommand{\numberMails}{2,987,591}
\newcommand{\numberPulls}{47,303}
\newcommand{\sizeTotal}{1.2 Terabyte}
\newcommand{\sizeNoMetrics}{40 Gigabyte}
\newcommand{\sizeStorageTotal}{440 Gigabyte}
\newcommand{\sizeStorageNoMetrics}{16 Gigabyte}

\begin{abstract}
The SmartSHARK repository mining data is a collection of rich and detailed information about the evolution of software projects. The data is unique in its diversity and contains detailed information about each change, issue tracking data, continuous integration data, as well as pull request and code review data. Moreover, the data does not contain only raw data scraped from repositories, but also annotations in form of labels determined through a combination of manual analysis and heuristics, as well as links between the different parts of the data set. The SmartSHARK data set provides a rich source of data that enables us to explore research questions that require data from different sources and/or longitudinal data over time.
\end{abstract}


\section{Introduction}

In the last years, we invested a large amount of effort to create a versatile data set about the evolution of software projects that combines data from different sources based on our SmartSHARK platform for replicable and reproducible software repository mining~\cite{Trautsch2017, Trautsch2020}. The core of this approach was to combine all data we generated for different publications in a single database, that grows with every publication. This does not only mean that we add more projects over time, but also that the amount of information for the projects already within the database increases. By now, our database contains the following data:

\begin{itemize}
    \item Data collected from Git, e.g., the commit messages, authors, dates, as well as the changed hunks. The clone of the Git repository at the time of collection is also stored to enable further analysis of the source code. 
    \item Data about the source code \textbf{for each commit} focused on Java, e.g., software metrics (size, complexity, documentation, code clones), static analysis warnings from PMD\footnote{https://pmd.github.io/}, and the number of nodes of each type in the AST of a file. 
    \item Data about code changes, i.e., the detection of change types with ChangeDistiller~\cite{Fluri2007}, as well as refactorings with RefDiff~\cite{Silva2017} and RefactoringMiner~\cite{Tsantalis2018}. 
    \item Data collected from Jira, i.e., the issues, comments, and changes to issues made. 
    \item Data collected from GitHub, i.e., issues, pull requests, and code reviews as part of pull requests. 
    \item Data collected from mailing lists, i.e., all emails from the developer and user mailing lists.
    \item Links  commits and issues, as well as links between commits and pull requests.
    \item Manually validated links between commits and bug issues, as well as the type of issues labeled as bug for 38 projects~\cite{Herbold2019}. 
    \item Manually validated line labels that mark which changes contributed to a bug fix for 23 projects as well as partial data for five additional projects~\cite{herbold2020largescale}.
    \item Annotations for commits and changes, i.e., bug fixing changes including their probable inducing changes, if changes modified Javadocs or inline comments, whether TODOs were added or removed, if test code changed or if we were able to detect refactorings. 
    \item Travis CI build logs and build status information for all projects that use Travis CI.
\end{itemize}

The identities of developers are managed in a separate collection that is not shared publicly, unless specifically requested with a description of the purpose. Hence, developers are only identified by the (random) object identifier in the database. %However, these are only the personal data contained in the metadata (e.g., account information), we did not pre-process text (e.g., commit messages, issue descriptions) to remove personal data that was possibly mentioned.

\section{Data Description}

This publication describes version 2.1 of the data, which is publicly available.\footnote{Full data: \url{http://141.5.100.155/smartshark_2_1.agz}\\Small version without code entity states, code group states, and clone instances: \url{http://141.5.100.155/smartshark_2_1.agz}\\Please check \url{https://smartshark.github.io/dbreleases/} for mirrors or newer releases.\\DOIs follow with official publication.} Older releases are available on our homepage, where we will also post future releases.\footnote{\url{https://smartshark.github.io/dbreleases/}} A description on how to setup the data for local use, as well as an example for accessing the data with Python is available online.\footnote{https://smartshark.github.io/fordevs/}

In the following, we describe the data sources, the tools we used for the data collection, the size and format of the data, the schema of our database, the sampling strategy we used and the list the projects for which data is available. 

\subsection{Data Sources}

The raw data was collected from four different sources. 

\begin{itemize}
\item Version control data is collected directly from a clone of the Git repository. The repositories are retrieved from GitHub.\footnote{\url{https://www.github.com/}}
\item Issue tracking data is collected from the Apache Jira\footnote{\url{https://issues.apache.org/jira/}} and GitHub. 
\item Pull request data is collected from GitHub. 
\item Continuous integration data is collected from Travis CI.\footnote{\url{https://www.travis-ci.com/}} 
\end{itemize}

All data is publicly available, but the tool vendors may require the registration to scrape the data.

\subsection{Data Collection Tools}

\begin{figure}
\includegraphics[width=\linewidth]{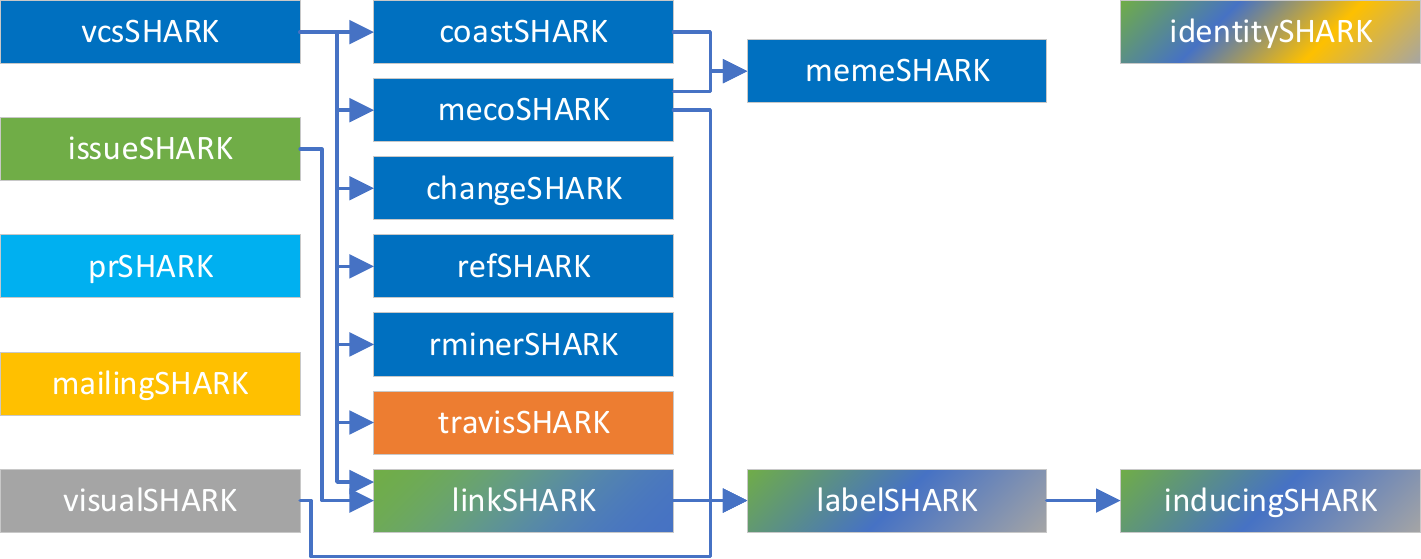}
\caption{Overview of data collection tools. The arrows indicate dependencies between tools. The colors indicate that different data sources are used (blue: Git repository; green: Jira and GitHub Issues; light blue: GitHub pull requests; yellow: mailing list archive; orange: Travis CI; grey: manual validation). A mix of colors means that data from different sources is required, as indicated by the dependencies and the colors. }
\label{fig:tools}
\end{figure}

Figure~\ref{fig:tools} shows the data collection tools we used. All tools are available on GitHub.\footnote{\url{https://github.com/smartshark/}} The vcsSHARK downloads a clone of the Git repository and collects metadata about commits. The coastSHARK, mecoSHARK, changeSHARK, refSHARK, and rminerSHARK use the clone of the repository to collect software metrics and detect refactorings. The memeSHARK removes duplicate software metrics and reduces the data volume. The travisSHARK collects data from Travis and links it to the commits. 
The prSHARK collects pull requests including reviews and comments from GitHub and links them to commits.
The mailingSHARK collects E-Mails from mailing lists.
The issueSHARK collects issue tracking data from Jira and GitHub issues. The linkSHARK establishes links between the issues and commits. The labelSHARK uses these links, the textual differences of changes, and changes to code metrics to compute labels for the commits. These labels are used by the inducingSHARK to find the probable changes that are inducing for the labels, e.g., for bugs. 

The visualSHARK is used for manual validation of data, e.g., of links between commits and issues, issue types, and changes that contribute to bug fixes. This information is used by the labelSHARK and inducingSHARK to improve data that relies this information, e.g., bug labels for commits. For completeness, we also mention the identitySHARK, which can be used to merge multiple identities of the same person in our data (e.g., different user name, same email). However, this data is not part of our public data set and will only be made available upon request if the desired usage is clearly specified and does not raise any ethical or data privacy related concerns. 

\subsection{Size and Format}

The data set currently contains \numberProjects{} projects, the manual validations are available for a subset of 38 projects. Overall, these projects have \numberCommits{} commits, \numberIssues{} issues, \numberPulls{} pull requests, and \numberMails{} emails. 
All data is stored in a MongoDB. The size of the complete MongoDB is \sizeTotal{}. This size drops drastically to about \sizeNoMetrics{}, if we omit the collections with code clone data and software metrics. 
The data is still growing and additional projects will also be made available through subsequent releases of the data set. 

Drivers for MongoDB are available for many programming languages.\footnote{https://docs.mongodb.com/drivers/} Additionally, we provide Object-Relational Mapping (ORM) for Python and Java. 

\subsection{Overview of the Database Schema}

\begin{figure*}
\centering
\includegraphics[width=\textwidth]{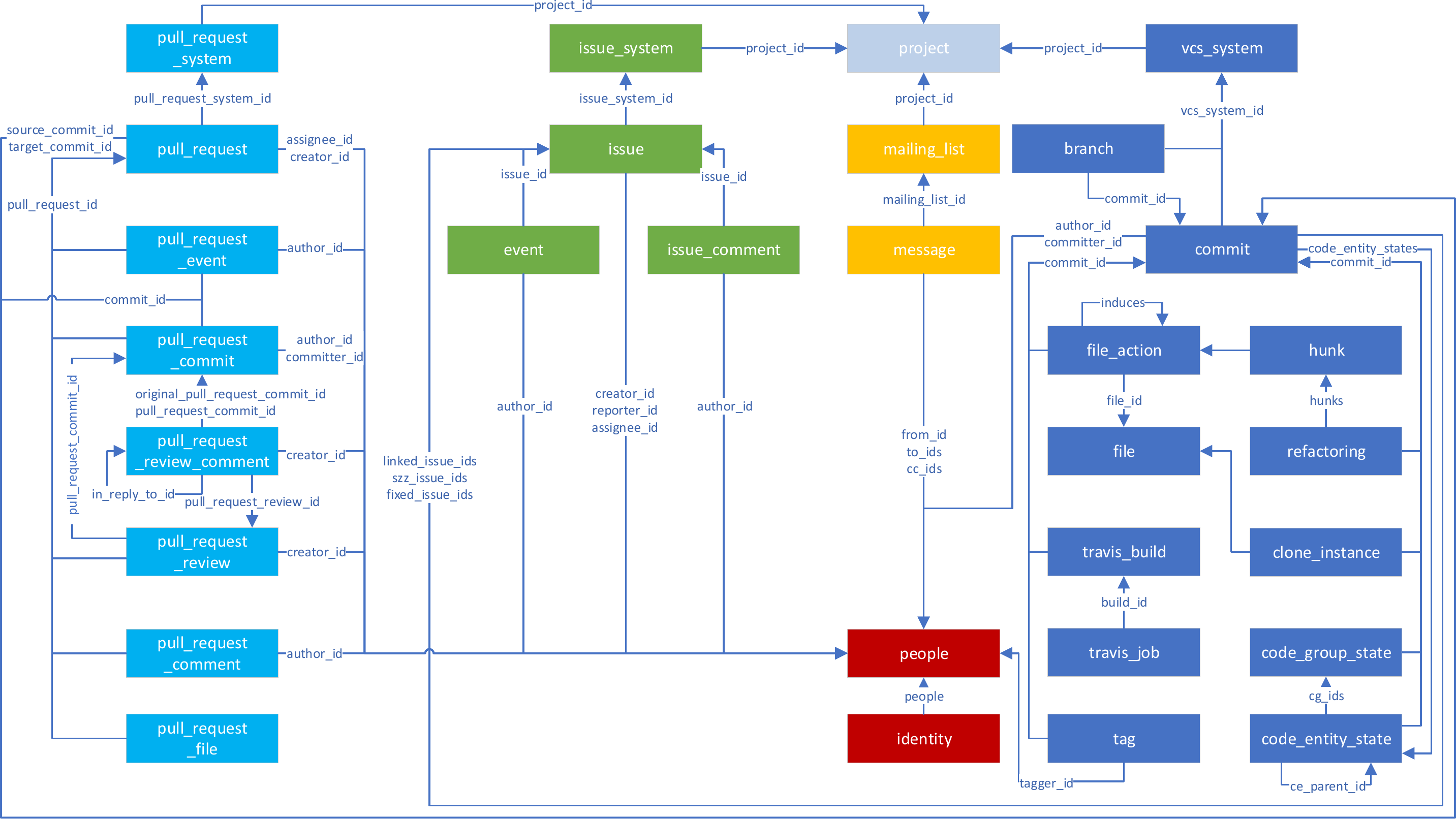}
\caption{Overview of the database schema and the relationships between the collections.}
\label{fig:schema}
\end{figure*}

We currently have data of four types of repositories: version control systems, issue tracking systems, pull request systems, and mailing lists. Figure~\ref{fig:schema} gives an overview of our database schema. A complete documentation is available online.\footnote{https://smartshark2.informatik.uni-goettingen.de/documentation/} Each project has an entry with the name and id. The software repositories are assigned to projects by their id. 

The simplest system are the mailing lists. The emails for the mailing lists are stored in the \texttt{message} collection. The issue tracking data is stored in three collections: \texttt{issue} stores the data about the issue itself, e.g., the title, description, and type; \texttt{issue\_comment} stores the discussion of the issue; and \texttt{event} stores any update made to the issue, e.g., changes of the type, status, or description. This way, the complete life-cycle including all updates is available in the database. The pull requests are organized similarly, but require seven collections, due to the direct relationship to source code and associated code reviews: \texttt{pull\_request} stores the metadata of the pull request, e.g., the title, description, and the associated branches; \texttt{pull\_request\_comment} stores the discussion of the pull request; \texttt{pull\_request\_event} stores any update made to the pull request; \texttt{pull\_request\_file} and \texttt{pull\_request\_commit} store references to files and commits within pull requests; and \texttt{pull\_request\_review} and \texttt{pull\_request\_review\_comment} store the information about code reviews. 

The version control system data is relatively complex, due to the diversity of the data stored. The main collection is \texttt{commit}, which contains the general metadata about the commits, e.g., the author, committer, revision hash, commit message, and time stamp. Moreover, \texttt{commit} also contains computed data, e.g., labels like bug fixing or links to issues. The \texttt{file\_action} group all changes made to a file in a commit, \texttt{hunk} contains the actual changes, including the diffs. The general information about the history is completed by the \texttt{branch} and \texttt{tag} collections. The \texttt{code\_group\_state} and \texttt{code\_entity\_state} contain the results of the static analysis we run on the repository at each commit. Code groups are, e.g., packages, code entities, are, e.g., files, classes, and methods. We removed duplicate code entities, e.g., files where the measurements did not change from one commit to the next. This way, we can reduce the data volume by over 11 Terabyte. To still allow the identification of the code entity states at the time of a commit, the \texttt{commit} collection contains a list to the correct code entities. While the code entities also contain a link to the commit for which they were measured, this link should be avoided, because users may inadvertently assume that they could find all code entities for a specific commits this way, which is not the case. The \texttt{clone\_instance} collection stores data about code clones. The automatically detected refactorings are stored in the \texttt{refactoring} collection. The \texttt{travis\_build} collection contains the general information about the build, e.g., time stamps and the build status, and the \texttt{travis\_job} collection contains the logs of the individual build jobs. 

The \texttt{people collection} is not associated with any specific data source. Instead, we map all metadata that contains accounts, names, or email addresses to this collection and store the name, email address, and user name. The identity collection contains list of people, that very likely belong to the same identity. We use our own identity merging algorithm, which is available online.\footnote{\url{https://github.com/smartshark/identitySHARK} (a scientific publication about our algorithm is not yet available)}

\subsection{Sampling Strategy and Representativness}

The data contains only projects from the Apache Software Foundation that have Java as the main language. The projects all have between 1,000 and 20,000 commits, i.e., the data does not contain very small or very large projects. The reason for the exclusion of very large projects is the data volume and processing time for the static analysis of each commit. 

While the sample is not randomly drawn, it should be representative for well-maintained Java projects that have high standards for their development processes, especially with respect to issue tracking. Moreover, the projects cover different kinds of applications, including build systems (ant-ivy), Web applications (e.g., jspwiki), database frameworks (e.g., calcite), big data processing tools (e.g., kylin), and general purpose libraries (commons). 

\subsection{List of Projects}

We have collected the data we described above for the following projects. Manually validated data is available for the italic projects. Travis CI data is available for all bold-faced projects. 

activemq, \textit{ant-ivy}, \textit{archiva}, \textbf{bigtop}, \textbf{\textit{calcite}}, \textbf{\textit{cayenne}}, \textbf{\textit{commons-bcel}}, \textbf{\textit{commons-beanutils}}, \textbf{\textit{commons-codec}}, \textbf{\textit{commons-collections}}, \textbf{\textit{commons-compress}}, \textbf{\textit{commons-configuration}}, \textbf{\textit{commons-dbcp}}, \textbf{\textit{commons-digester}}, \textbf{commons-imaging}, \textbf{\textit{commons-io}}, \textbf{\textit{commons-jcs}}, \textbf{\textit{commons-jexl}}, \textbf{\textit{commons-lang}}, \textbf{\textit{commons-math}}, \textbf{\textit{commons-net}}, \textbf{commons-rdf}, \textit{commons-scxml}, \textbf{\textit{commons-validator}}, \textbf{\textit{commons-vfs}}, \textbf{curator}, cxf-fediz, \textit{deltaspike}, derby, directory-fortress-core, directory-kerby, directory-studio, \textit{eagle}, falcon, \textbf{fineract}, \textbf{flume}, \textbf{freemarker}, \textit{giraph}, \textit{gora}, helix, \textbf{httpcomponents-client}, \textbf{httpcomponents-core}, jackrabbit, jena, \textit{jspwiki}, kafka, \textbf{\textit{knox}}, \textbf{\textit{kylin}}, \textit{lens}, \textbf{\textit{mahout}}, \textbf{\textit{manifoldcf}}, maven, mina-sshd, \textbf{nifi}, \textit{nutch}, oozie, openjpa, openwebbeans, \textbf{\textit{opennlp}}, \textbf{\textit{parquet-mr}}, \textbf{pdfbox}, phoenix, pig, \textbf{ranger}, roller, \textbf{samza}, \textit{santuario-java}, \textbf{storm}, \textbf{streams}, \textbf{struts}, \textit{systemml}, \textbf{tez}, \textit{tika}, \textit{wss4j}, xerces2-j, \textbf{xmlgraphics-batik}, \textbf{zeppelin}, zookeeper.

The bug inducing changes are not available for maven, because the project uses multiple issue trackers, which we currently cannot handle. \footnote{The release 2.2 of this data set with more projects and this missing inducing data is scheduled for December 2022. This preprint will then be updated with the final list of projects available for the challenge.} 

\section{Usage Examples}

The SmartSHARK data is versatile and allows different kinds of research. In the past, we have focused mostly on the analysis of bugs, as well as longitudinal analysis of trends within the development history. Below, we list some examples of papers that used (a subset of) this data set. Please note that some of papers below are still under review and not yet published in their final versions.

\begin{itemize}
    \item We evaluated defect prediction data quality with a focus on SZZ and manual validation~\cite{Herbold2019}. The manuscript describes how we manually validated the links between commits and issues, as well as the issue types and how we used SmartSHARK to create release-level defect prediction data.
    \item We evaluated trends of static analysis warnings from PMD and usage of custom rules for PMD as well as the impact on defect density~\cite{Trautsch2020b}.
    \item We evaluated the impact of static source code metrics and static analysis warnings from PMD on just-in-time defect prediction~\cite{Trautsch2020a}.
    \item We used the manually validated issue type data to train and evaluate issue type prediction models~\cite{Herbold2020a}.
    \item We provided the data for the modelling of the developer behaviour through Hidden Markov Models (HMMs)~\cite{Herbold2019a}.
    \item We analyzed the tangling within bug fixing commits as well as the capability of researchers to manually identify tangling~\cite{Herbold2020}.
    \item We conducted an initial evaluation of a cost model for defect prediction~\cite{Herbold2019c}. 
\end{itemize}

\section{Possible Research Questions}

% Code improvements due to code reviews in pull requests (nur wenns auch verlinkungen gibt)
% 

The strength of our data is the capability to reason over data from different information sources. Questions regarding differences between the discussions on mailing lists and within issue trackers can be answered without scraping data from multiple sources. Moreover, the static analysis results and the labeling of changes allow for research into the relationship, e.g., between refactorings and self-admitted technical debt or bug fixes. The availability of manually validated data enables us to evaluate the validity of heuristics, as well as the development of improvements of heuristics, e.g., for the labeling of bug fixes. Moreover, while we already established many links between the data sources, there are more opportunities that could be considered, e.g., the links between the mailing list and commits, or the mailing list and reported issues. Similarly, the links between pull requests and bugs can be explored, e.g., to understand why post release bugs were not spotted during code review. 

\section{Limitations}

The greatest limitation of the SmartSHARK data is the number of projects for which data is available. The reason for this is the large computational effort required for the static analysis of the Java code of each commit. This not only limits the external validity due to the sample size, but also due to a focus on Java as programming language. In the future, we plan to overcome this limitation by extending the database with a large set of projects, for which we omit the static analysis and, thereby, are able to scale up the number of projects. While this will not support the same research questions, there are many interesting questions that can be answered without a static analysis of the source code for each commit. 

\section{Conclusion}

The SmartSHARK data set provides a rich source of data that enables us to explore research questions that require data from different sources and/or longitudinal data over time. Since all data is stored in a single data base, results are easy to reproduce. The data is still growing and future releases will further extend the data with more projects and additional data sources. 

\nocite{*}
\bibliographystyle{IEEEtran}
\bibliography{IEEEabrv,./literature}

\begin{thebibliography}{10}
\providecommand{\url}[1]{#1}
\csname url@samestyle\endcsname
\providecommand{\newblock}{\relax}
\providecommand{\bibinfo}[2]{#2}
\providecommand{\BIBentrySTDinterwordspacing}{\spaceskip=0pt\relax}
\providecommand{\BIBentryALTinterwordstretchfactor}{4}
\providecommand{\BIBentryALTinterwordspacing}{\spaceskip=\fontdimen2\font plus
\BIBentryALTinterwordstretchfactor\fontdimen3\font minus
  \fontdimen4\font\relax}
\providecommand{\BIBforeignlanguage}[2]{{%
\expandafter\ifx\csname l@#1\endcsname\relax
\typeout{** WARNING: IEEEtran.bst: No hyphenation pattern has been}%
\typeout{** loaded for the language `#1'. Using the pattern for}%
\typeout{** the default language instead.}%
\else
\language=\csname l@#1\endcsname
\fi
#2}}
\providecommand{\BIBdecl}{\relax}
\BIBdecl

\bibitem{Trautsch2017}
F.~Trautsch, S.~Herbold, P.~Makedonski, and J.~Grabowski, ``{Addressing
  problems with replicability and validity of repository mining studies through
  a smart data platform},'' \emph{Empirical Software Engineering}, Aug. 2017.

\bibitem{Trautsch2020}
A.~Trautsch, F.~Trautsch, S.~Herbold, B.~Ledel, and J.~Grabowski, ``The
  smartshark ecosystem for software repository mining,'' in \emph{Proc. of the
  2020 Int. Conf. Softw. Eng. - Demonstrations Track}, 2020.

\bibitem{Fluri2007}
B.~Fluri, M.~W{\"u}rsch, M.~Pinzger, and H.~Gall, ``Change distilling: Tree
  differencing for fine-grained source code change extraction,'' \emph{IEEE
  Transactions on Software Engineering}, vol.~33, no.~11, pp. 725--743, 2007.

\bibitem{Silva2017}
D.~{Silva} and M.~T. {Valente}, ``Refdiff: Detecting refactorings in version
  histories,'' in \emph{2017 IEEE/ACM 14th International Conference on Mining
  Software Repositories (MSR)}, May 2017, pp. 269--279.

\bibitem{Tsantalis2018}
\BIBentryALTinterwordspacing
N.~Tsantalis, M.~Mansouri, L.~M. Eshkevari, D.~Mazinanian, and D.~Dig,
  ``Accurate and efficient refactoring detection in commit history,'' in
  \emph{Proceedings of the 40th International Conference on Software
  Engineering}, ser. ICSE '18.\hskip 1em plus 0.5em minus 0.4em\relax New York,
  NY, USA: ACM, 2018, pp. 483--494. [Online]. Available:
  \url{http://doi.acm.org/10.1145/3180155.3180206}
\BIBentrySTDinterwordspacing

\bibitem{Herbold2019}
S.~Herbold, A.~Trautsch, and F.~Trautsch, ``Issues with szz: An empirical
  assessment of the state of practice of defect prediction data collection,''
  2019.

\bibitem{herbold2020largescale}
S.~Herbold, A.~Trautsch, B.~Ledel, A.~Aghamohammadi, T.~A. Ghaleb, K.~K.
  Chahal, T.~Bossenmaier, B.~Nagaria, P.~Makedonski, M.~N. Ahmadabadi,
  K.~Szabados, H.~Spieker, M.~Madeja, N.~Hoy, V.~Lenarduzzi, S.~Wang,
  G.~Rodríguez-Pérez, R.~Colomo-Palacios, R.~Verdecchia, P.~Singh, Y.~Qin,
  D.~Chakroborti, W.~Davis, V.~Walunj, H.~Wu, D.~Marcilio, O.~Alam, A.~Aldaeej,
  I.~Amit, B.~Turhan, S.~Eismann, A.-K. Wickert, I.~Malavolta, M.~Sulir,
  F.~Fard, A.~Z. Henley, S.~Kourtzanidis, E.~Tuzun, C.~Treude, S.~M. Shamasbi,
  I.~Pashchenko, M.~Wyrich, J.~Davis, A.~Serebrenik, E.~Albrecht, E.~U. Aktas,
  D.~Strüber, and J.~Erbel, ``Large-scale manual validation of bug fixing
  commits: A fine-grained analysis of tangling,'' 2020.

\bibitem{Trautsch2020b}
\BIBentryALTinterwordspacing
A.~Trautsch, S.~Herbold, and J.~Grabowski, ``{A Longitudinal Study of Static
  Analysis Warning Evolution and the Effects of PMD on Software Quality in
  Apache Open Source Projects},'' \emph{Empirical Software Engineering}, 2020.
  [Online]. Available: \url{https://doi.org/10.1007/s10664-020-09880-1}
\BIBentrySTDinterwordspacing

\bibitem{Trautsch2020a}
------, ``Static source code metrics and static analysis warnings for
  fine-grained just-in-time defect prediction,'' in \emph{Proc. of the 2020
  Int. Conf. Softw. Maintenance and Evolution}, 2020.

\bibitem{Herbold2020a}
S.~Herbold, A.~Trautsch, and F.~Trautsch, ``On the feasibility of automated
  issue type prediction,'' https://arxiv.org/abs/2003.05357, 2020.

\bibitem{Herbold2019a}
\BIBentryALTinterwordspacing
V.~Herbold, ``Mining developer dynamics for agent-based simulation of software
  evolution,'' Dissertation, University of Goettingen, Germany, 2019. [Online].
  Available: \url{http://hdl.handle.net/21.11130/00-1735-0000-0003-C15C-C}
\BIBentrySTDinterwordspacing

\bibitem{Herbold2020}
\BIBentryALTinterwordspacing
S.~Herbold, A.~Trautsch, and B.~Ledel, ``Large-scale manual validation of
  bugfixing changes,'' Mar 2020. [Online]. Available: \url{osf.io/acnwk}
\BIBentrySTDinterwordspacing

\bibitem{Herbold2019c}
S.~Herbold, ``On the costs and profit of software defect prediction,''
  \emph{IEEE Transactions on Software Engineering}, no.~01, pp. 1--1, dec 2019.

\bibitem{Zhao2017}
\BIBentryALTinterwordspacing
Y.~Zhao, H.~Leung, Y.~Yang, Y.~Zhou, and B.~Xu, ``Towards an understanding of
  change types in bug fixing code,'' \emph{Information and Software
  Technology}, vol.~86, pp. 37 -- 53, 2017. [Online]. Available:
  \url{http://www.sciencedirect.com/science/article/pii/S0950584917301313}
\BIBentrySTDinterwordspacing

\bibitem{Sliwerski2005}
\BIBentryALTinterwordspacing
J.~\'{S}liwerski, T.~Zimmermann, and A.~Zeller, ``When do changes induce
  fixes?'' in \emph{Proceedings of the 2005 International Workshop on Mining
  Software Repositories}, ser. MSR '05.\hskip 1em plus 0.5em minus 0.4em\relax
  New York, NY, USA: ACM, 2005, pp. 1--5. [Online]. Available:
  \url{http://doi.acm.org/10.1145/1082983.1083147}
\BIBentrySTDinterwordspacing

\bibitem{Spadini2018}
\BIBentryALTinterwordspacing
D.~Spadini, M.~Aniche, and A.~Bacchelli, ``{PyDriller: Python framework for
  mining software repositories},'' in \emph{Proceedings of the 2018 26th ACM
  Joint Meeting on European Software Engineering Conference and Symposium on
  the Foundations of Software Engineering - ESEC/FSE 2018}.\hskip 1em plus
  0.5em minus 0.4em\relax New York, New York, USA: ACM Press, 2018, pp.
  908--911. [Online]. Available:
  \url{http://dl.acm.org/citation.cfm?doid=3236024.3264598}
\BIBentrySTDinterwordspacing

\bibitem{Herzig2013}
\BIBentryALTinterwordspacing
K.~Herzig, S.~Just, and A.~Zeller, ``It's not a bug, it's a feature: How
  misclassification impacts bug prediction,'' in \emph{Proceedings of the 2013
  International Conference on Software Engineering}, ser. ICSE '13.\hskip 1em
  plus 0.5em minus 0.4em\relax Piscataway, NJ, USA: IEEE Press, 2013, pp.
  392--401. [Online]. Available:
  \url{http://dl.acm.org/citation.cfm?id=2486788.2486840}
\BIBentrySTDinterwordspacing

\bibitem{Buse2010}
\BIBentryALTinterwordspacing
R.~P.~L. Buse and W.~R. Weimer, ``Learning a metric for code readability,''
  \emph{IEEE Trans. Softw. Eng.}, vol.~36, no.~4, pp. 546--558, Jul. 2010.
  [Online]. Available: \url{http://dx.doi.org/10.1109/TSE.2009.70}
\BIBentrySTDinterwordspacing

\bibitem{Scalabrino2018}
S.~Scalabrino, M.~Linares-Vásquez, R.~Oliveto, and D.~Poshyvanyk, ``A
  comprehensive model for code readability,'' \emph{Journal of Software:
  Evolution and Process}, vol.~30, no.~6, p. e1958, 2018.

\bibitem{Gousios2013}
G.~Gousios, ``The ghtorrent dataset and tool suite,'' in \emph{Proceedings of
  the 10th Working Conference on Mining Software Repositories}, ser. MSR
  '13.\hskip 1em plus 0.5em minus 0.4em\relax Piscataway, NJ, USA: IEEE Press,
  2013, pp. 233--236.

\bibitem{Trautsch2019}
\BIBentryALTinterwordspacing
F.~Trautsch, S.~Herbold, and J.~Grabowski, ``Are unit and integration test
  definitions still valid for modern java projects? an empirical study on
  open-source projects,'' \emph{Journal of Systems and Software}, vol. 159, p.
  110421, 2020. [Online]. Available:
  \url{http://www.sciencedirect.com/science/article/pii/S0164121219301955}
\BIBentrySTDinterwordspacing

\bibitem{Trautsch2019a}
A.~{Trautsch}, ``Effects of automated static analysis tools: A multidimensional
  view on quality evolution,'' in \emph{2019 IEEE/ACM 41st International
  Conference on Software Engineering: Companion Proceedings (ICSE-Companion)},
  May 2019, pp. 184--185.

\bibitem{Trautsch2016}
\BIBentryALTinterwordspacing
F.~Trautsch, S.~Herbold, P.~Makedonski, and J.~Grabowski, ``Adressing problems
  with external validity of repository mining studies through a smart data
  platform,'' in \emph{Proceedings of the 13th International Conference on
  Mining Software Repositories}, ser. MSR ’16.\hskip 1em plus 0.5em minus
  0.4em\relax New York, NY, USA: Association for Computing Machinery, 2016, p.
  97–108. [Online]. Available: \url{https://doi.org/10.1145/2901739.2901753}
\BIBentrySTDinterwordspacing

\end{thebibliography}

\end{document}